\documentclass[12pt]{article}
\usepackage{graphicx}
\usepackage{amsfonts}
\usepackage{amssymb}
\usepackage{amsmath}
\usepackage{bm}

\begin{document}

\centerline{\Large\bf Gravity-wave detectors as probes of extra
dimensions}
\bigskip
\bigskip
\centerline{{\bf Chris Clarkson and Roy Maartens}\footnote{email:
chris.clarkson@port.ac.uk, roy.maartens@port.ac.uk }}
\bigskip
\bigskip
\smallskip
\centerline{Institute of Cosmology \& Gravitation, University of
Portsmouth,}
\centerline{ Portsmouth~PO1 2EG, UK}
\bigskip

\begin{abstract}

If string theory is correct, then our observable Universe may be a
3-dimensional ``brane" embedded in a higher-dimensional spacetime.
This theoretical scenario should be tested via the
state-of-the-art in gravitational experiments -- the current and
upcoming gravity-wave detectors. Indeed, the existence of extra
dimensions leads to oscillations that leave a spectroscopic
signature in the gravity-wave signal from black holes. The
detectors that have been designed to confirm Einstein's prediction
of gravity waves, can in principle also provide tests and
constraints on string theory.

\end{abstract}
\newpage

Black holes are fascinating objects that are crucial to our
theoretical understanding of gravity. They also provide a testing
ground for gravitational theory, because they probe the
strong-gravity regime. The deepest access to this regime is
provided by gravity waves, which should be detected by the current
and upcoming generation of experiments. These experiments open up
a new opportunity to test candidate quantum gravity theories.
Amongst these, string theory makes the radical prediction that
spacetime has extra spatial dimensions -- so that gravity
propagates in higher dimensions and has extra polarizations. These
polarizations would leave a ``smoking gun" imprint in the
gravity-wave signal from black hole events.

How do the extra spatial dimensions escape detection and
observation at low energies? Recent developments in string theory
provide a revolutionary new way to achieve this -- the brane-world
scenario, in which the observable Universe, including Standard
Model fields, is confined to a 3-dimensional ``brane", while
gravity propagates in the full ``bulk"
spacetime~\cite{bworld,rs,rev}. The extra dimensions can be large
(relative to the Planck length), and as a consequence, the true
fundamental energy scale of gravity can be much lower than the
4-dimensional Planck scale, perhaps even down to the TeV-level. In
that event, small black holes would be produced in the upcoming
generation of particle colliders~\cite{collbh}. TeV-scale gravity
would bring quantum gravity within reach of the laboratory. But if
the fundamental scale is much higher (while still lower than the
4-dimensional Planck scale), then laboratory testing is pushed
further into the future. By contrast, gravity-wave detectors
provide access to black hole events at energies high enough to
carry any detectable signatures of extra-dimensional gravity.

The brane-world scenario encompasses a very wide variety,
including string theory solutions and phenomenological models. In
order to investigate perturbations of black holes that have formed
via gravitational collapse, one needs a model which is simple
enough to have an exact background solution, but rich enough to
include key aspects of astrophysical black holes and of string
theory. Perhaps the best candidate is the black string
solution~\cite{chr}, in which the 4-dimensional Schwarzschild
metric is embedded in a 5-dimensional Randall-Sundrum type
model~\cite{rs}. One important feature of the Randall-Sundrum
scenario is that the self-gravity of the brane is included, which
leads to a curvature (``warping") of the extra dimension.

\begin{figure}[!t]
\begin{center}
\includegraphics[width=3in]{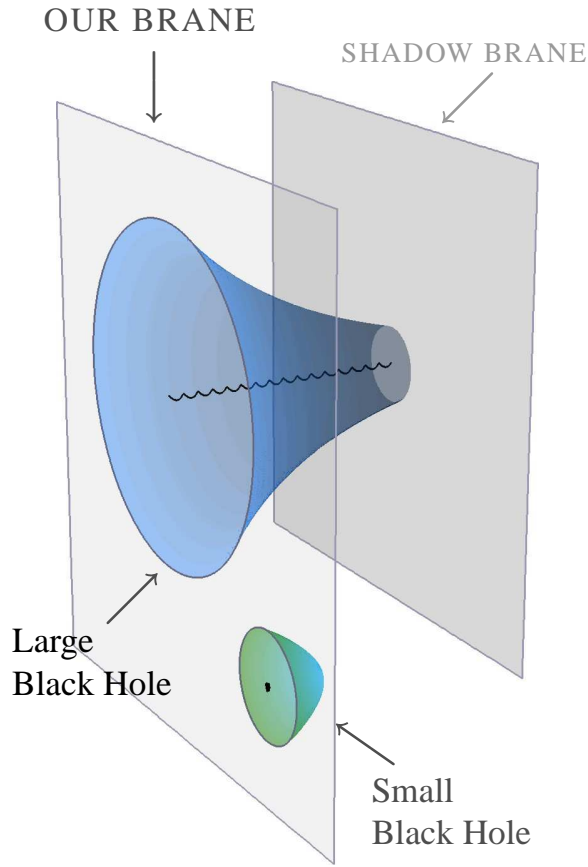}
\caption{\small Schematic of brane-world black holes. (The brane
separation is greatly exaggerated.) }\label{F1}
\end{center}
\end{figure}

In the original Randall-Sundrum model, two Minkowski branes, with
equal and opposite tensions (vacuum energies) enclose a patch of
5-dimensional anti de Sitter spacetime, with curvature scale
$\ell$. The bulk metric satisfies the 5-dimensional Einstein
equations, $G_{ab}=6g_{ab}/\ell^2$, and the boundary conditions at
the branes are the Israel junction conditions. There is a mirror
symmetry at each brane (a feature of some string theory
solutions). The black string metric satisfies the same equations
in the bulk, and is given by
\begin{eqnarray} \label{black string metric}
ds^2 & = & {\rm e}^{-2|y|/\ell} \left[ -(1 -2GM/r)dt^2+{dr^2\over
1 -2GM/r}+r^2\,d\Omega^2 \right] + dy^2\,.
\end{eqnarray}
The induced metric on the``visible" brane at $y=0$ is the
Schwarzschild metric. If the horizon radius on the brane is much
greater than the size of the extra dimension, $2GM\gg d$, then we
can think of the extension of the horizon into the 5th dimension
as being cut off by the ``shadow" brane at $y=d$~\cite{ks}. By
contrast, small black holes, $2GM \ll d$, do not ``see" the brane
and behave like 5-dimensional black holes, localized on the
visible brane. No exact solution is known for a black hole
localized on the brane, but the nature of small localized black
holes has been investigated numerically~\cite{ktn}. This set-up is
illustrated in Fig.~\ref{F1}.

\begin{figure}[!t]
\begin{center}
\includegraphics[width=.8\textwidth]{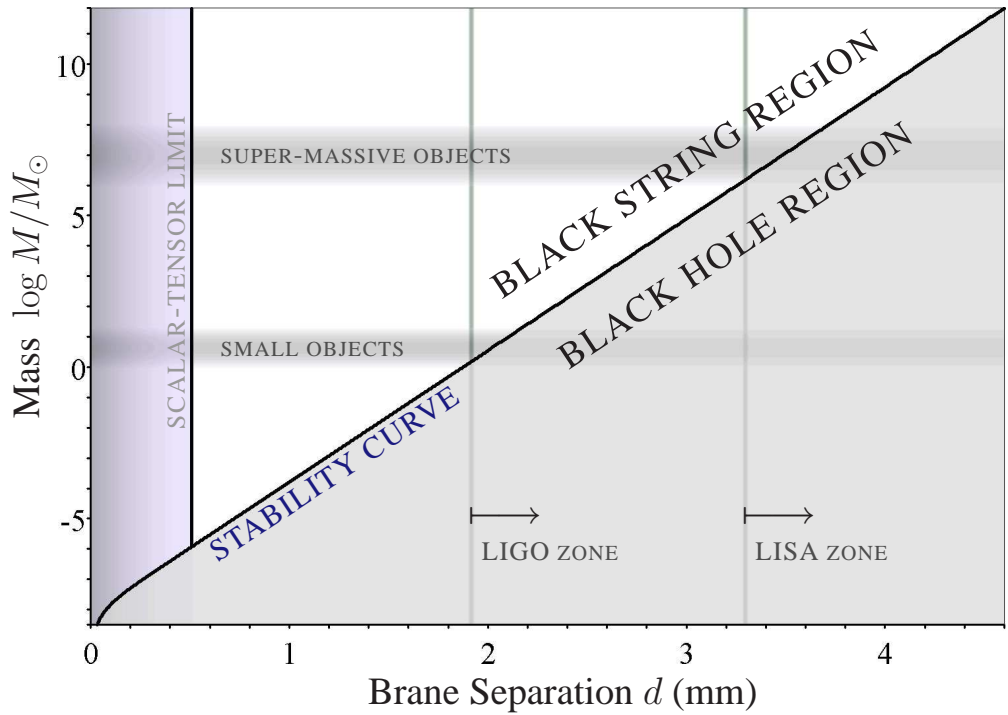}\\
\caption{\small The parameter space of the black string, for
$\ell=0.1$mm. The stability curve is approximated by
Eq.~(\ref{gl}). The LIGO and LISA zones follow from
Eq.~(\ref{freq}). }\label{F2}
\end{center}
\end{figure}

The shadow brane simultaneously provides an infrared cut-off to
shut down the Gregory-Laflamme long-wavelength instability of the
black string. The condition for stability is~\cite{gl}
\begin{equation}\label{gl}
GM \gtrsim 0.1\, \ell \,{\rm e}^{d/\ell}\,.
\end{equation}
In fact small black holes ($GM < 0.1 \ell {\rm e}^{d/\ell}$)
localized on the brane can be seen as products of the instability,
since there is evidence for a black string to black hole
transition as the outcome of the instability~\cite{cho}. The
parameters are also constrained by solar system observations and
by laboratory tests of Newton's law, via a remarkable interlinking
of different gravitational properties. The shadow brane must be
far enough away that its gravitational influence on the visible
brane is within observational limits. The brane separation is a
massless scalar degree of freedom (the ``radion") on the visible
brane, so that the effective theory on the visible brane is of
Brans-Dicke type~\cite{gt}, with $\omega_{\rm bd}=3( {\rm
e}^{2d/\ell}-1)/2$. Solar system observations impose the lower
limit~\cite{bit,wy} $\omega_{\rm bd} \gtrsim 4 \times 10^4$, so
that $d/\ell\gtrsim 5$. The allowed region in parameter space is
shown in Fig.~\ref{F2}. Table-top tests of Newton's law impose the
constraint~\cite{exp} $\ell\lesssim 0.1\,$mm, and we use this
upper limit in Fig.~\ref{F2}.

We also show in Fig.~\ref{F2} the detectable minimum brane
separation, corresponding to a maximum characteristic frequency,
for the LIGO and LISA detectors. The extra-dimensional
polarizations of the graviton are realized on the visible brane as
effectively massive Kaluza-Klein modes, which produce late-time
oscillations in the gravity-wave signal, as described below.
Because of the boundary conditions at the branes, the Kaluza-Klein
masses form a discrete tower~\cite{rs}
\begin{equation}
m_n=(z_n/\ell){\rm e}^{-d/\ell}~\mbox{where}~ J_1(z_n)\approx
0\,,~ n=1,2,3,\cdots
\end{equation}
The corresponding tower of frequencies for the oscillating massive
modes is~\cite{scm}
\begin{equation}\label{freq}
f_n = z_n {\rm e}^{26.9-d/\ell}
(0.1\,\text{mm}/\ell)\,\,\text{Hz.}
\end{equation}
The $z_n$'s have a spacing of order one. It is striking that,
unlike 4-dimensional black hole quasinormal modes, these
frequencies are {\em independent} of the mass $M$. For $\ell =
0.1\,$mm, the lowest frequency $f_1$ is within the LIGO upper
limit $f_{\rm max}\sim 10^4\,$Hz  for $d /\ell\sim 20$. The LISA
upper limit gives $d /\ell\sim 30$. Typical optimal frequencies
are
\begin{eqnarray}
&&\text{LIGO:}~~f_1 \sim 100\,{\rm Hz}~\text{which
implies}~M>100M_\odot\,,\nonumber
\\
&&\text{LISA:}~~\,f_1 \sim 0.01\,{\rm Hz}~\text{which
implies}~M>10^6M_\odot\,. \nonumber
\end{eqnarray}
Figure~\ref{F2} also shows that the detectors can in principle see
a black string instability event for small or supermassive
objects, which would be one of the best sources of massive modes.

Current detectors fall into two categories: those that are very
nearly up-and-running, such as LIGO, TAMA and GEO, detect high
frequencies around 1kHz, while proposed space detectors such as
LISA are designed for low frequencies around the mHz range. All
are designed to cover two or three orders of magnitude around
these values. Will they be able to detect massive modes of
higher-dimensional gravitons, if such modes exist? The smallest
$d/\ell$ these detectors can measure is illustrated by the
vertical bands in Fig.~\ref{F2} for high- and low-frequency
detectors. But these detectors can do rather better than just the
$f_1$ mode: the tower of frequencies~(\ref{freq}) is dense enough
to allow about ten discrete mass modes in each decade. Thus all
brane separations to the right of the vertical bands will have a
range of around 30 massive mode frequencies detectable by present
technology.

The gravity-wave output on the brane from black string events is
in general a combination of the massless zero-mode, $m=0$, which
reproduces the standard 4-dimensional signal (a damped
single-frequency quasinormal ringing followed by a power-law
tail~\cite{nol}), and the massive Kaluza-Klein modes, with much
less damping and late-time oscillations~\cite{scm}. The details
follow from solving the wave equation for perturbations of the
metric~(\ref{black string metric}), which reduces to five coupled
linear wave equations on the brane, corresponding to the 5
polarizations of the 5-dimensional graviton (and generalizing the
2-dimensional system of the Regge-Wheeler and Zerilli
equations~\cite{nol}).

\begin{figure}[!t]
\begin{center}
\includegraphics[width=0.6\textwidth]{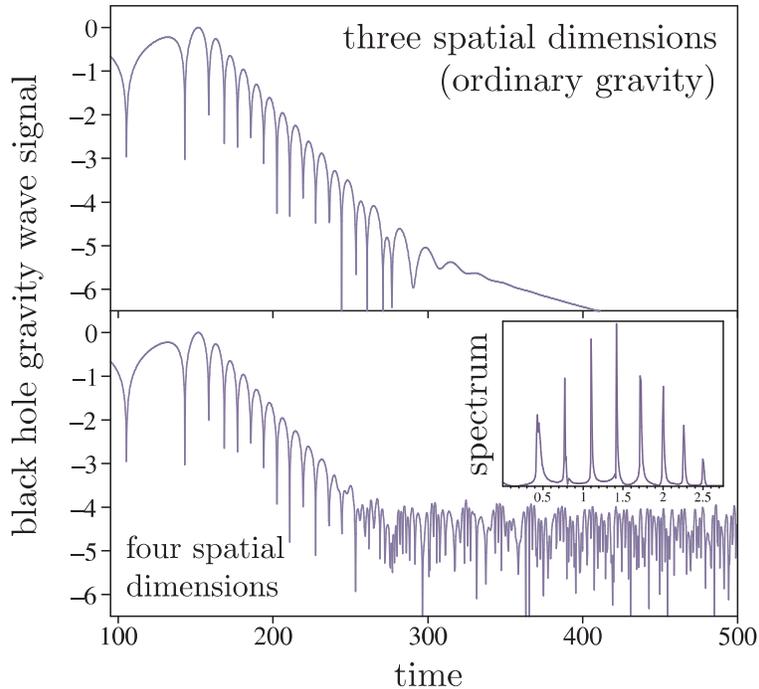}
\caption{\small Gravity-wave signal from a black hole in General
Relativity (upper panel) and a brane-world black string (lower
panel). In the black string case, the early-time waveform is
dominated by the zero-mode (the General Relativity ringdown), but
at late times the massive modes take over and remove the General
Relativity power-law tail. The Fourier transform of the late-time
signal is shown in the inset. }\label{F3}
\end{center}
\end{figure}

The results of simulations~\cite{scm} are shown in Fig.~\ref{F3},
for the $L=2$ axial modes (in a generalized Regge-Wheeler gauge),
in the case of initial data that weakly excites the lowest massive
modes -- for example, an encounter between a heavy black string
and a small localized black hole or other body on the brane. In
General Relativity the late-time signal is featureless~-- a simple
power-law tail. For the black string however, an observer would
see a high-frequency signal that persists at late times. The
Fourier transform of this late-time signal shows {\em a
spectroscopic series of spikes, located at the characteristic
frequencies of the massive modes}.

Initial data that strongly excites the massive modes corresponds
to events which ``mainly take place in the bulk", such as the
merger of two black strings. In this case there is very similar
late-time behaviour, but the early-time waveform has no General
Relativity-like ringdown. In all cases, the key common feature is
that \emph{the signal is made of discrete frequencies which are
independent of the mass of the black string}.

Can the massive signal be resolved by gravity-wave detectors? This
entails both the frequency of the massive modes, discussed above,
and their relative amplitude. The relative amplitude of the
massive to massless signal depends strongly on whether the initial
data has primary support on the brane or in the bulk. For the
brane-based perturbation in Fig.~\ref{F3}, the detection of the
tail would require a signal-to-noise ratio $\gtrsim 10^4$, just
beyond the projected capabilities of LISA~\cite{wy}. The relative
massive amplitude tends to increase with decreasing $d/\ell$.
However, small $d/\ell$ implies large $f_n$, so the best prospect
of seeing this type of signal lies with high frequency detectors.
Such constraints do not apply to bulk-based initial data, like
black string mergers. Since the zero-mode energy is small in these
situations, we expect the strength of the massive mode
oscillations to be comparable to the quasinormal ringing amplitude
in the analogous 4-dimensional case.

At late times and large distances on the brane from the black
string, the massive modes decay very slowly, as $t^{-5/6}$~--
unlike the exponential decay of the zero mode. Other than direct
observation of an event, what consequences could this gentle decay
have? A black string which is being continually excited will
steadily emit gravity waves, typically with some excitation of the
massive modes. For example, for two black strings rotating around
each other in the latter stages of merger, massive modes would be
continually produced. An observer would see a signal composed of a
mixture of a General Relativity inspiral waveform, interlaced with
a massive signal. After merger the massive modes would persist for
much longer than the massless mode~-- for a massive mode, the
signal strength drops by a factor of 10 after $\sim16$ times the
lifetime of the massless mode. Typically there would be many
events like this taking place in a galaxy over time. As the signal
from one merger never properly decays away, massive modes of the
higher-dimensional graviton would accumulate over time, reaching
an equilibrium strength proportional to the number of mergers.
This would lead to {\em an integrated massive mode contribution to
the stochastic gravity-wave background.} The gravity-wave
background would in principle give access to the massive mode
frequencies.

Furthermore, since the massive modes travel below light-speed,
there will be potentially observable {\em time-delays} in their
arrival, which could be of order seconds or longer for distant
sources.

The discrete nature of the late-time Fourier transform of the
black string waveform is the most important observable feature.
Its detection would provide clear evidence of extra dimensions,
and could further give the direct spectroscopic measurement of the
Kaluza-Klein masses $m_n$. This in turn provides information about
the size and shape of the extra dimension -- gravity-wave
detectors can in principle explore the geography of large extra
dimensions.

These results are based on a specific simple model of brane-world
black holes (the only exact solution with reasonable astrophysical
properties that we know of). But we expect that qualitatively
similar features will arise for other models with large extra
dimensions, since they all have a discrete tower of massive
Kaluza-Klein modes. Could it be possible that we will be able to
observe or exclude predictions of string theory using current
technology? We might not have to wait long to find out.

\[ \]
{\bf Acknowledgements}

CC and RM are supported by PPARC. We thank Sanjeev Seahra for his
crucial collaboration in the work on which this essay is based,
and for ongoing discussions of the subject.


\newpage

\begin{thebibliography}{99}

\bibitem{bworld}
P. Horava and E. Witten, Nucl. Phys. {\bf B460}, 506 (1996)
[hep-th/9510209];\\ A. Lukas, B. A. Ovrut, K. S. Stelle and D.
Waldram, Phys. Rev. D{\bf 59}, 086001 (1999)
[hep-th/9803235];\\
N. Arkani-Hamed, S. Dimopoulos and G. Dvali, Phys. Lett. {\bf
B429}, 263 (1998) [hep-ph/9803315].

\bibitem{rs}
L. Randall and R. Sundrum, Phys. Rev. Lett. {\bf 83}, 3370 (1999)
[hep-ph/9905221].

\bibitem{rev}
For a review, see e.g., R. Maartens, Liv. Rev. Rel. {\bf 7}, 7
(2004) [gr-qc/0312059].

\bibitem{collbh}
For a review, see e.g., P. Kanti, Int. J. Mod. Phys. {\bf A19},
4899 (2004) [hep-ph/0402168].

\bibitem{chr}
A. Chamblin, S. W. Hawking and H. S. Reall, Phys. Rev. D{\bf 61},
065007 (2000) [hep-th/9909205].

\bibitem{ks}
S. Kanno and J. Soda, Class. Quant. Grav. {\bf 21}, 1915 (2004)
[gr-qc/0311074].

\bibitem{ktn}
H. Kudoh, T. Tanaka and T. Nakamura, Phys. Rev. D{\bf 68}, 024035
(2003) [gr-qc/0301089].

\bibitem{gl}
R. Gregory, Class. Quant. Grav. {\bf 17}, L125 (2000)
[hep-th/0004101].

\bibitem{cho}
M.~W.~Choptuik, L.~Lehner, I.~Olabarrieta, R.~Petryk, F.~Pretorius
and H.~Villegas,
Phys.\ Rev.\ D{\bf 68}, 044001 (2003) [arXiv:gr-qc/0304085];\\
B. Kol, E. Sorkin and T. Piran, Phys. Rev. D{\bf 69}, 064031
(2004) [hep-th/0309190];\\ H. Kudoh and T. Wiseman,
hep-th/0409111;\\ B. Kol, hep-th/0411240.

\bibitem{gt}
J. Garriga and T. Tanaka, Phys. Rev. Lett. {\bf 84}, 2778 (2000)
[hep-th/9911055];\\
T. Wiseman, Class. Quant. Grav. {\bf 19}, 3083 (2002)
[hep-th/0201127].

\bibitem{bit}
B. Bertotti, L. Iess and P. Tortora, Nature {\bf 425}, 374 (2003).

\bibitem{wy}
C. M. Will and N. Yunes, Class. Quant. Grav. {\bf 21}, 4367 (2004)
[ gr-qc/0403100.].

\bibitem{exp}
J.C. Long et al., Nature {\bf 421}, 922 (2003) [hep-ph/0210004].

\bibitem{scm}
S. S. Seahra, C. Clarkson and R. Maartens, Phys. Rev. Lett., {\bf
94} 121302 (2005) [gr-qc/0408032].

\bibitem{nol}
For a review, see e.g., H. Nollert, Class. Quant. Grav. {\bf 16},
R159 (1999).


\end{thebibliography}
\end{document}